# P-wave meson properties with Wilson quarks[*]

M. Wingate[a], T. DeGrand[a], S. Collins[b], U. Heller[b]

[a]Physics Department, University of Colorado,
Boulder, CO 80309

[b]SCRI, Florida State University,
Tallahassee, FL 32306

We describe two calculations involving P-wave mesons made of Wilson quarks: the strong coupling constant $\alpha_s$ in the presence of two flavors of light dynamical fermions and the mass and decay constant of the $a_1$ meson.

## 1. INTRODUCTION

P-wave meson spectroscopy has become increasingly interesting to the lattice community in the past few years, mainly for heavy quark systems. The spectroscopy of heavy and light quark systems has its own intrinsic charm, and additionally the S-P mass splitting in quarkonium can be used to determine the QCD strong coupling constant through a technique pioneered by the Fermilab group[1], where it is used to fix the lattice spacing and the strong coupling constant at a particular $q$ value is extracted from the expectation value of the plaquette.

## 2. THE STRONG COUPLING CONSTANT

We measure the strong coupling constant $\alpha_s$ from simulations using the HEMCGC $\beta = 5.6$ configurations with two flavors of light dynamical fermions[2], and from quenched simulations at $\beta = 6.0$ generated as part of the MILC collaboration program[3]. Our particles have valence Wilson quarks, and so our study complements the work of the NRQCD collaboration[4], who used the same dynamical fermion simulations but non-relativistic heavy quarks.

One begins by using the S-P mass difference of a heavy quark system to set the lattice spacing. In our case we have so far concentrated on the difference

$$\Delta M_{^3P_1-\overline{S}} = M(^3P_1) - \frac{1}{4}\left[3M(^3S_1) + M(^1S_0)\right]$$
$$= 442 \text{ MeV} \quad (1)$$

where the $^3P_1$ meson is the axial vector meson created by the interpolating field $\bar{\psi}\gamma_i\gamma_5\psi$. We have also reconstructed masses of the full lattice multiplet corresponding to the P-wave system. This method allows us to compute

$$\Delta M_{^1P_1-\overline{S}} = 458 \text{ MeV} \quad (2)$$

and

$$\Delta M_{\overline{P}-\overline{S}} = 457 \text{ MeV} \quad (3)$$

in addition to (1). Comparing lattice spacings obtained by fixing each of the above quantities gives a lattice spacing $1/a = 1900(50)(100)$ MeV, where the first error is statistical and the second is systematic.

The coupling constant is defined through the plaquette in the so-called $\alpha_V$ scheme,

$$-\ln\langle\frac{1}{3}TrU_P\rangle = \frac{4\pi}{3}\alpha_V(3.41/a)\times$$
$$\left[1 - (1.19 + 0.07n_f)\alpha_V + \mathcal{O}(\alpha_V^2)\right]. \quad (4)$$

Here $\frac{1}{3}TrU_P = 0.56500(2)$ so $\alpha_V(3.41/a) = 0.179(10)$, where the error is dominated entirely by $\mathcal{O}(\alpha_V^3)$ effects. This is for two flavors of dynamical quarks. To convert to the physically interesting case (for charmonium) of three flavors, we repeat the calculation on the quenched

---

$\beta = 6.0$ data set, where $1/a = 2290(270)$ MeV and $\alpha_V(3.41/a) = 0.152(4)$, then run one data set to the same $Q$ as the other and extrapolate linearly in $1/\alpha_V$ from $N_f = 0$ and 2 to $N_f = 3$: we find $\alpha_V^{(n_f=3)}(7.81\ \text{GeV}) = 0.178(15)$. The largest contributions to the error come from two sources: the uncertainty in the lattice spacing and the $\mathcal{O}(\alpha_V^3)$ uncertainty in (4).

Finally, to run to the Z mass (where numbers are conventionally compared) we convert from $\alpha_V$ to $\alpha_{\overline{MS}}$ using

$$\alpha_{\overline{MS}}(Q) = \alpha_V(Qe^{5/6})\left[1 + 2\alpha_V/\pi + \mathcal{O}(\alpha_V^2)\right] \quad (5)$$

and then run in $Q$ using the formulas of Rodrigo and Santamaria[5]. Our result is

$$\alpha_{\overline{MS}}^{(n_f=5)}(M_Z) = 0.108(6). \quad (6)$$

A picture displaying our results with others at the Z mass is shown in Fig. 1.

One puzzling feature of the calculation is the very different lattice spacings found by us using heavy Wilson quarks and the lattice spacing of the NRQCD group. The numbers are shown in Table 1, along with all other lattice spacings extracted from this data set of which we are aware: in the table S and W label staggered and Wilson valence quarks, the "force" is from the string tension[6], and the zero quark mass line is from an extrapolation when it is available. Clearly scaling violations are large in this data set. They are expected to be of order $a$ for the hadron spectroscopy, which presumably for this reason does not agree with experiment, and of order $a^2$ for the "force" and NRQCD.

## 3. THE $a_1$ MESON

The $a_1$ meson is the isovector axial vector meson made of nonstrange quarks. Long ago even the mass of the $a_1$ was controversial, but nowadays the $a_1$ is most readily studied via tau decay. The coupling of an $a_1$ (of polarization $\epsilon_i$) to the W-boson is parameterized[7] by the $a_1$ decay constant, $f_{a_1}$, by

$$\langle 0|\bar{\psi}\gamma_i\gamma_5\psi|a_1\rangle = f_{a_1}\epsilon_i. \quad (7)$$

To calculate the mass and decay constant of the $a_1$ is a straightforward lattice exercise (provided,

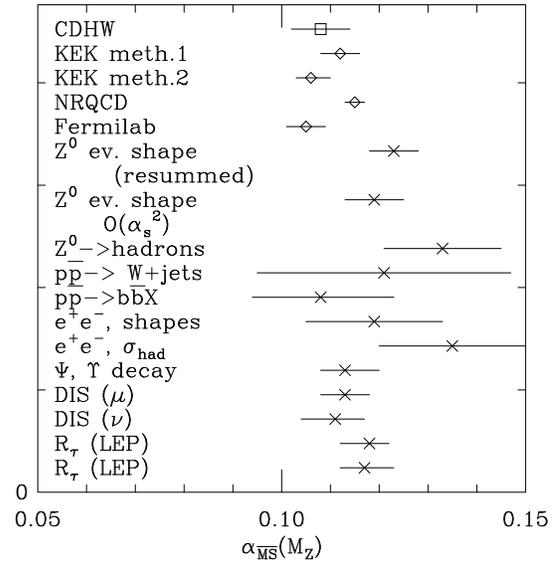

Figure 1. Several measurements of the strong coupling constant at the Z-mass. The square marks our result. Diamonds represent other lattice calculations, and crosses are experimental results.

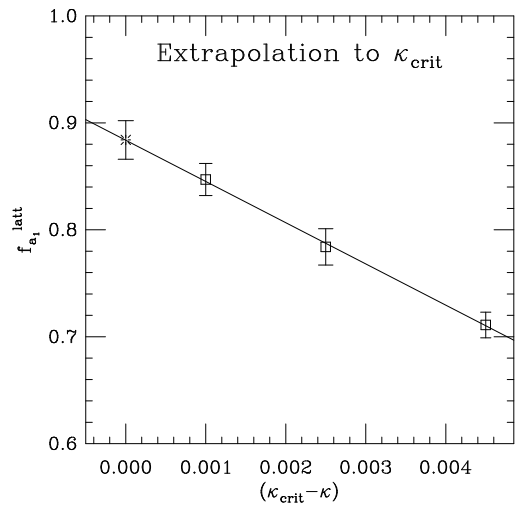

Figure 2. The squares indicate measurements of $f_{a_1}^{latt}$ for three light $\kappa$ values. The asterisk marks the extrapolation to $\kappa_{\text{crit}} = 0.1610$.

Table 1
Inverse lattice spacings (in MeV) from various observables from the dynamical fermion data sets. Statistical errors are 50 – 100 MeV.

| $am_q$ | "force" | $m_\rho$(W) | $m_p$(W) | $m_\rho$(S) | $m_P$(S) | NRQCD |
|---|---|---|---|---|---|---|
| 0.025 | 1935 | 2000 | 1685 | — | — | — |
| 0.010 | 2055 | 2140 | 1800 | — | — | 2400 |
| 0.0 | 2135 | 2230 | 1875 | 1800 | 1660 | — |

of course that one can see a signal!) We merely carried the propagator calculations of Sec. 2 down to light quark mass. We used a combination of Coulomb-gauge Gaussian shell model sources and sinks and the axial current itself as a sink, and extracted the mass and decay constants from correlated fits to the two propagators. We performed a jacknife extrapolation of the mass and decay constants into $\kappa_c$, (see Fig. 2).

We have so far only done this exercise for the $\beta = 5.6$ data sets. We have three values of light quark mass. We used our S-P mass splittings to set the lattice spacing. To convert the lattice decay constant to a continuum one we used the tadpole improvement scheme of Lepage and Mackenzie[8]:

$$\langle 0|O^{cont}(\mu = 1/a)|a_1\rangle_{\overline{MS}} = \langle 0|O^{latt}(a)|a_1\rangle \times$$
$$f(m)a^{-2}(am_{a_1})^2(1 + A_0\alpha_s + \mathcal{O}(\alpha_s^2))$$
$$+ \mathcal{O}(a) \quad (8)$$

where $f(m) = 1/4$ (at $\kappa = \kappa_c$) converts the lattice field renormalization to the continuum, $A_0 = 0.31$ in tadpole improvement, and the factor $a^{-2}$ converts the dimensionless lattice number to its continuum result. The coupling constant is defined through the plaquette and is run down to a scale $\mathcal{O}(1/a)$.

We did not see any dependence of our measured quantities on quark mass, and so we present only averages here. We find $m_{a_1} = 1270(70)$ MeV (expt. 1230(40)) and $f_{a_1} = 0.31(3)$ GeV$^2$ (expt. 0.25(2)). Comparing with experiment is actually rather nontrivial because of the large width of the $a_1$; final state interactions are important. The "experimental" number for the decay constant comes from a phenomenological analysis by Isgur, Morningstar, and Reader[9], which uses experimental input which is probably now obsolescent.

## 4. CONCLUSIONS

We plan to perform both of these measurements on the HEMCGC $\beta = 5.3$ dynamical Wilson fermion configurations, and to complete a $\beta = 6.0$ quenched data set. These measurements will allow us to search for lattice spacing systematics as well as possible effects of dynamical fermions.

This work was supported by the U.S. Department of Energy and by the National Science Foundation. Simulations were performed on the CM-2 at the Supercomputer Computations Research Institute and on the Paragon at the San Diego Supercomputer Center.